\documentclass[acmsmall,screen,nonacm]{acmart}
\usepackage{booktabs,tabularx,array}
\usepackage{amsmath}
\usepackage{needspace}
\usepackage{tikz}
\usetikzlibrary{arrows.meta,positioning,fit,backgrounds}
\definecolor{aisnavy}{HTML}{536170}
\definecolor{aisblue}{HTML}{E8EDF3}
\definecolor{aispurple}{HTML}{E8EDF3}
\definecolor{aisgreen}{HTML}{F0F1F3}
\definecolor{aisgold}{HTML}{F0F1F3}
\definecolor{aisgray}{HTML}{F0F1F3}
\definecolor{aisrose}{HTML}{E1E4E8}
\tikzset{
 aisbox/.style={draw=aisnavy,fill=aisblue,line width=.45pt,align=center,inner sep=4pt},
 aispanel/.style={draw=aisnavy,dashed,line width=.45pt},
 aisflow/.style={->,draw=black!85,line width=.45pt},
 aisnote/.style={font=\scriptsize,align=center},
 pics/aisdocument/.style={code={
   \draw[draw=aisnavy,fill=white,line width=.5pt] (-.16,-.21)--(.16,-.21)--(.16,.10)--(.05,.21)--(-.16,.21)--cycle;
   \draw[draw=aisnavy,line width=.45pt] (.05,.21)--(.05,.10)--(.16,.10);
   \draw[draw=aisnavy,line width=.45pt] (-.10,.04)--(.09,.04) (-.10,-.04)--(.09,-.04) (-.10,-.12)--(.03,-.12);
 }},
 pics/aisuser/.style={code={
   \draw[draw=aisnavy,fill=aisgold,line width=.5pt] (0,.11) circle (.10);
   \draw[draw=aisnavy,fill=aisgold,line width=.5pt,rounded corners=2pt] (-.16,-.23) rectangle (.16,-.03);
 }},
 pics/aisagent/.style={code={
   \draw[draw=aisnavy,line width=.5pt] (0,.13)--(0,.23);
   \fill[aisnavy] (0,.23) circle (.025);
   \draw[draw=aisnavy,fill=aispurple,line width=.5pt,rounded corners=1pt] (-.18,-.16) rectangle (.18,.13);
   \fill[aisnavy] (-.08,.03) circle (.025) (.08,.03) circle (.025);
   \draw[draw=aisnavy,line width=.5pt] (-.06,-.08)--(.06,-.08);
 }},
 pics/aisshield/.style={code={
   \draw[draw=aisnavy,fill=aisgreen,line width=.5pt] (-.18,.21)--(.18,.21)--(.15,-.07)--(0,-.23)--(-.15,-.07)--cycle;
   \draw[draw=aisnavy,line width=.7pt] (-.09,.01)--(-.02,-.07)--(.10,.09);
 }},
 pics/aisserver/.style={code={
   \foreach \y in {-.12,.10}{
     \draw[draw=aisnavy,fill=aisgray,line width=.5pt] (-.20,\y-.08) rectangle (.20,\y+.08);
     \fill[aisnavy] (-.12,\y) circle (.023);
     \draw[draw=aisnavy,line width=.5pt] (-.02,\y)--(.13,\y);
   }
 }}
}

\setcopyright{none}
\newcolumntype{Y}{>{\raggedright\arraybackslash}X}
\definecolor{pathblue}{RGB}{32,78,114}
\definecolor{pathgold}{RGB}{137,88,32}
\begin{document}
\title[Security of Agent-Integrated Software]{Security of Agent-Integrated Software: When Human Operations and Agent Actions Coexist}
\author{Ding Yang}
\affiliation{\institution{State Key Laboratory for Novel Software Technology, Nanjing University}\city{Nanjing}\country{China}}
\email{dingyang@smail.nju.edu.cn}
\orcid{0009-0005-9684-2046}
\author{Yuchen Ling}
\affiliation{\institution{State Key Laboratory for Novel Software Technology, Nanjing University}\city{Nanjing}\country{China}}
\email{yuchenling@smail.nju.edu.cn}
\orcid{0009-0006-9227-3824}
\author{Shengcheng Yu}
\affiliation{\institution{Technical University of Munich}\city{Heilbronn}\country{Germany}}
\email{shengcheng.yu@tum.de}
\orcid{0000-0003-4640-8637}
\author{Zhenyu Chen}
\affiliation{\institution{State Key Laboratory for Novel Software Technology, Nanjing University}\city{Nanjing}\country{China}}
\email{zychen@nju.edu.cn}
\orcid{0000-0002-9592-7022}
\author{Chunrong Fang}
\affiliation{\institution{State Key Laboratory for Novel Software Technology, Nanjing University}\city{Nanjing}\country{China}}
\email{fangchunrong@nju.edu.cn}
\orcid{0000-0002-9930-7111}
\renewcommand{\shortauthors}{Yang et al.}
\begin{abstract}
Agent-Integrated Software (AIS) embeds an intelligent agent in a conventional application, supporting both human operations and agent actions.
Human operations let users make precise changes and inspect results, while agent actions carry out routine or multi-step tasks.
These complementary roles make coexistence a likely long-term feature of many software systems.
Human operations and agent actions affect the same software state and can use one another's results.
Therefore, security policies must remain effective across both paths.
We argue that AIS security must be assessed at the level of the whole software system.
Protecting the agent and the conventional software core separately does not establish that they are secure together.
To guide security analysis of AIS as a whole, we organize the problems arising from this coexistence into four categories: context misuse, authorization violation, execution control, and effect integrity.
Using these categories, we examine how current practices address the security problems in AIS and where their protection remains limited.
Building on this analysis, we identify research opportunities in preserving information provenance, enforcing policy across operation paths, maintaining valid authorization over time, and managing persistent effects and recovery.
This resulting perspective provides a conceptual framework for understanding and improving the security of AIS.
\end{abstract}
\maketitle
\section{Introduction}
\label{sec:intro}
Embedding an intelligent agent in conventional software gives users another way to use its functions~\cite{xie2026intent,vscode2026security}.
We use \emph{Agent-Integrated Software} (AIS) to describe a conventional application with a built-in intelligent agent.
AIS supports both \emph{human operations}, which users choose and perform themselves, and \emph{agent actions}, which the agent selects and executes toward user-specified goals.
Human operations let users make precise changes, inspect results, and handle work they choose not to delegate.
Agent actions can carry out routine or multi-step tasks without requiring users to specify every step~\cite{yao2023react}.
These complementary roles suggest that human operations and agent actions will coexist in many software systems over the long term~\cite{horvitz1999principles,bainbridge1983ironies}.
This \emph{coexistence} includes separate, alternating, and overlapping use of the two forms of operation.

The coexistence of human operations and agent actions matters for security because they access shared resources and can use one another's results.
Information supplied by users may contain instructions that redirect the agent's actions~\cite{greshake2023injection}.
Conversely, agent actions can produce results that users later edit or share.
A security rule intended to protect shared data can fail if the software enforces it for only one form of operation.
For example, if a report in a document management system must remain confidential, blocking human operations from sharing it is insufficient when the agent's sending tool can still send it without the required authorization.
This leads to our position: \emph{AIS security must be assessed at the level of the whole software system.}
Protecting the agent and the conventional software core separately does not establish that they are secure together.
The software must enforce its security rules across both human operations and agent actions.
Their permissions may differ, but each operation must remain within its authorized scope.

To support a systematic analysis of AIS security, we organize the problems arising from this coexistence into four categories.
\emph{Context misuse} occurs when the agent uses information in ways that the software's security rules do not allow.
It may mistake content supplied through human operations for instructions, or pass protected data to a model or tool that is not allowed to receive it.
\emph{Authorization violation} occurs when human operations or agent actions exceed their permissions.
An operation prohibited through one route must not become possible simply by switching to the other.
\emph{Execution control} concerns whether users can control an ongoing task and whether its actions remain authorized as conditions change.
When permissions change or users cancel a task, the software must prevent further actions that are no longer authorized.
\emph{Effect integrity} concerns whether the individual and combined effects of operations respect the software's security rules.
This includes protecting results during later use, correcting harmful changes without damaging other authorized work, and recording which actions produced those effects.

To describe how users express and control delegated tasks within this analysis, we use \emph{Intent-Level Interaction Abstraction} (IIA)~\cite{yu2026engineering}.
IIA describes how users express goals, review proposed actions, and understand outcomes.
It helps us distinguish what users ask the agent to achieve from what they authorize it to do, and identify which decisions remain under user control.
For example, a request to prepare a report does not by itself authorize publication.
This distinction helps us explain the authorization and consent problems above: the software must enforce the authorized scope when the agent turns a goal into concrete actions.
IIA provides a way to describe these interaction requirements; their enforcement remains the responsibility of the software.

Together, we build a conceptual framework for understanding AIS security when human operations and agent actions coexist.
We then use this framework to examine what current practices protect and where gaps remain, and derive a research agenda from those gaps.
Our focus is the security of AIS as a whole.
This perspective aims to help developers identify missing protections and make informed security decisions when integrating agents into existing software.
It also provides a starting point for researchers to investigate how these protections can remain effective as agent capabilities and software workflows evolve.
Section~\ref{sec:categories} develops the problem categories.
Section~\ref{sec:practices} examines current practices, and Section~\ref{sec:opportunities} presents research opportunities.
Section~\ref{sec:conclusion} concludes.

\section{Security Problems in Agent-Integrated Software}
\label{sec:categories}
The four security categories concern how software data influences agent decisions, how delegated tasks use authority, and how their effects persist as software state changes. Their interpretation depends on the system boundary and security assumptions defined below.

\subsection{System scope and security assumptions}
\label{sec:coexistence}
\emph{Agent-Integrated Software integrates a conventional application core with a built-in intelligent agent, supporting both human operations and goal-directed agent actions.} We use this definition for systems that meet three conditions. Users can still perform operations through the software interface. An agent can select and execute operations toward a goal, and those actions use the software's resources under its security policies. These operations include reading protected information as well as changing data. A built-in agent need not run in the same process as the software core or hold unrestricted credentials. Table~\ref{tab:scope} gives boundary cases by configuration. ``LLM-integrated application'' can include a passive text generator, while a ``tool-using agent'' need not retain an application interface. Mixed-initiative systems describe allocation of initiative more broadly~\cite{horvitz1999principles}. Assistants and copilots also qualify as AIS when they meet these conditions. The AIS definition identifies the human operations, agent actions, resources, and policies that must be considered together in a security analysis. Shared-control task environments provide examples of this coexistence~\cite{huq2025cowpilot,shao2024collaborative,barres2025tau2}.

\begin{table}[t]
\caption{Scope of AIS in this perspective. The configuration determines whether a system is in scope.}
\label{tab:scope}
\small
\begin{tabularx}{\linewidth}{@{}>{\raggedright\arraybackslash}p{.38\linewidth}>{\raggedright\arraybackslash}p{.12\linewidth}Y@{}}
\toprule
\textbf{Configuration} & \textbf{In scope?} & \textbf{Reason} \\
\midrule\midrule
Application supporting human operations and an agent using scoped domain APIs & Yes & Both forms of operation use the software's resources and must follow its policies \\
\midrule
Integrated agent restricted to protected retrieval & Yes & Retrieval still requires controls over access and disclosure \\
\midrule
Chat widget that only generates text for the user to copy & No & No integrated agent execution of domain operations \\
\midrule
External automation controlling an unrelated application & Not alone & Tool use alone does not establish built-in integration. Assessment must specify the boundary of the composed software system \\
\midrule
Agent-only service, or fixed workflow without an intelligent agent & No & Lacks retained human operations or goal-directed agent actions, respectively \\
\bottomrule
\end{tabularx}
\end{table}

\begin{figure}[!t]
\centering
\includegraphics[width=\linewidth]{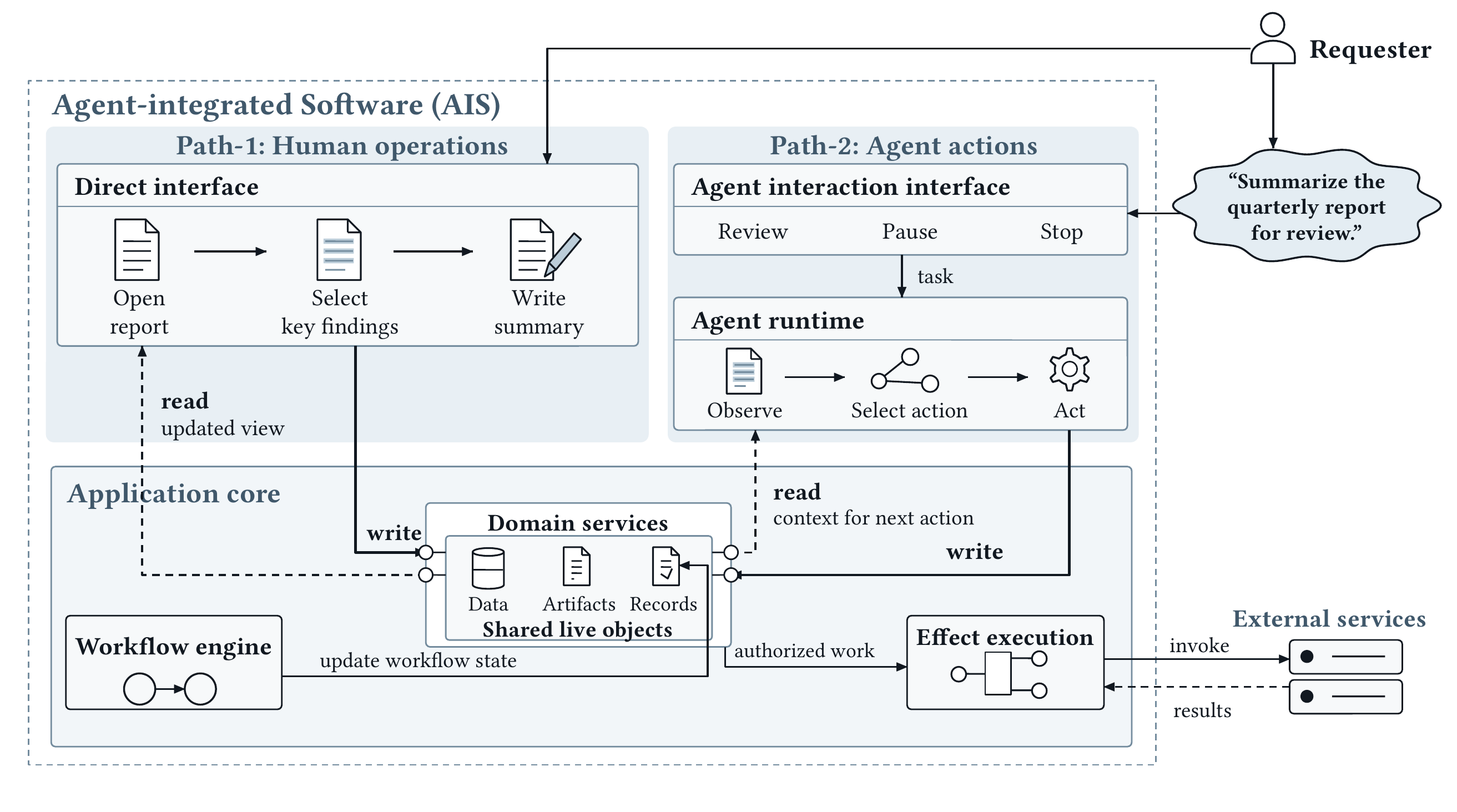}
\caption{AIS architecture: human operations and agent actions share software services and state. Solid arrows show operations and state changes; dashed arrows show observations.}
\Description{For the same summarization task, the direct route shows the user opening the report, selecting key findings, and writing a summary. On the delegated route, a requester supplies the illustrative natural-language goal, "Summarize the quarterly report for review," through a cloud-shaped dialogue bubble connected to the agent interaction interface. The direct interface and agent runtime read and write one shared object collection through the domain-services component. The upper path connects human writes to agent reads, while the lower path connects agent writes to human reads. The workflow engine has an arrow labeled "update workflow state" pointing to Records among the shared live objects. An authorized-work arrow from domain services reaches an effect-execution component, which invokes external services and receives results. All three components belong to the application core.}
\label{fig:architecture}
\end{figure}

Figure~\ref{fig:architecture} shows human operations and agent actions sharing software services and state. Human changes can alter the context of later agent steps, while agent updates can alter subsequent human views and operations. These influences pass through the application core. The workflow engine updates workflow state in the shared records, and the effect-execution component invokes external services. The components represent logical responsibilities. The software boundary does not imply that all components are trusted or that external completion is under its control. The person requesting a task, the account executing it, the resource owner, and the person approving an action may differ. A service credential identifies the account executing an operation but does not show that the requester authorized the task to use every privilege held by that account~\cite{lampson1992authentication}.

\paragraph{Using IIA to describe security requirements.}
\emph{Intent-Level Interaction Abstraction describes how users express, inspect, and redirect delegated work at the task level.} It covers the goal and relevant resources, proposed actions, decisions that require approval, user interventions, and known outcomes~\cite{yu2026engineering}. IIA helps express the requirement that actions permitted during execution must match the task and decisions presented to the user. For example, an interface may ask the user to approve one document, while execution later selects files from a changing folder. Recording that the user clicked ``approve'' does not resolve this mismatch. IIA describes the interaction without prescribing an agent implementation or a storage format. To enforce this requirement, the software can associate each task with its \emph{requester and execution identity}, \emph{goal}, \emph{resource and recipient scope}, \emph{allowed effects}, \emph{reserved decisions}, \emph{validity conditions}, and \emph{control rights}. The software needs this information to enforce policy, although the interface need not present it all as form fields. For a report task, the user may authorize preparation of version $v$ for group $g$, while publication requires a separate approval. That approval should identify both $v$ and the reviewed membership of $g$, and the requester should be able to revoke further submissions. Defaults can supply some of these limits, but users must be able to understand the limits that affect what the task may do.

Before allowing a protected operation, the software must identify the actual resources and recipients, authenticate the requester and execution identity, and check that resource permissions, task scope, and required approvals still permit the action. An approval must come from the authorized decision-maker. The agent cannot supply it by interpreting the request. Task records or capabilities can carry these restrictions, while service receipts provide evidence for the outcomes shown to the user. Established delegation mechanisms may already support these checks. IIA helps explain what they must enforce in terms of the task the user understands.

\paragraph{Assets, adversary, and trust.}
Protected assets include personal information and its copies in agent context, memory, and logs. They also include other confidential data, software objects and workflows, resources subject to quotas, and records of authorization and outcomes. We consider an adversary who can submit requests within ordinary account rights, edit content they control, or supply retrieved external content. These capabilities are specified per example and do not include arbitrary modification of trusted enforcement. Neither the planner nor the content it reads can grant authority. We trust the components that authenticate decisions and enforce permissions to perform their stated checks, and assume that providers honor their documented contracts without assuming additional guarantees. Legitimate users also remain subject to policy. Interpretation errors, delays, and recovery failures become security concerns when they violate the protections defined for the task.

\paragraph{Operations, effects, and evidence.}
An \emph{operation} is an attempted action in the software, and its \emph{request} proposes execution. \emph{Admission} is acceptance by an identified service. An \emph{effect} is a resulting disclosure, state change, or resource use. \emph{Commitment} is the point at which the service considers a specified effect to have taken place. The observed \emph{outcome} is what receipts or other evidence confirm. An operation may produce several effects before failing. A service accepting a request does not necessarily mean that the effect has occurred, or that the requester knows whether it has occurred.

\subsection{Basis for the four security categories}
\label{sec:derivation}
\begin{figure}[!t]
\centering
\includegraphics[width=\linewidth]{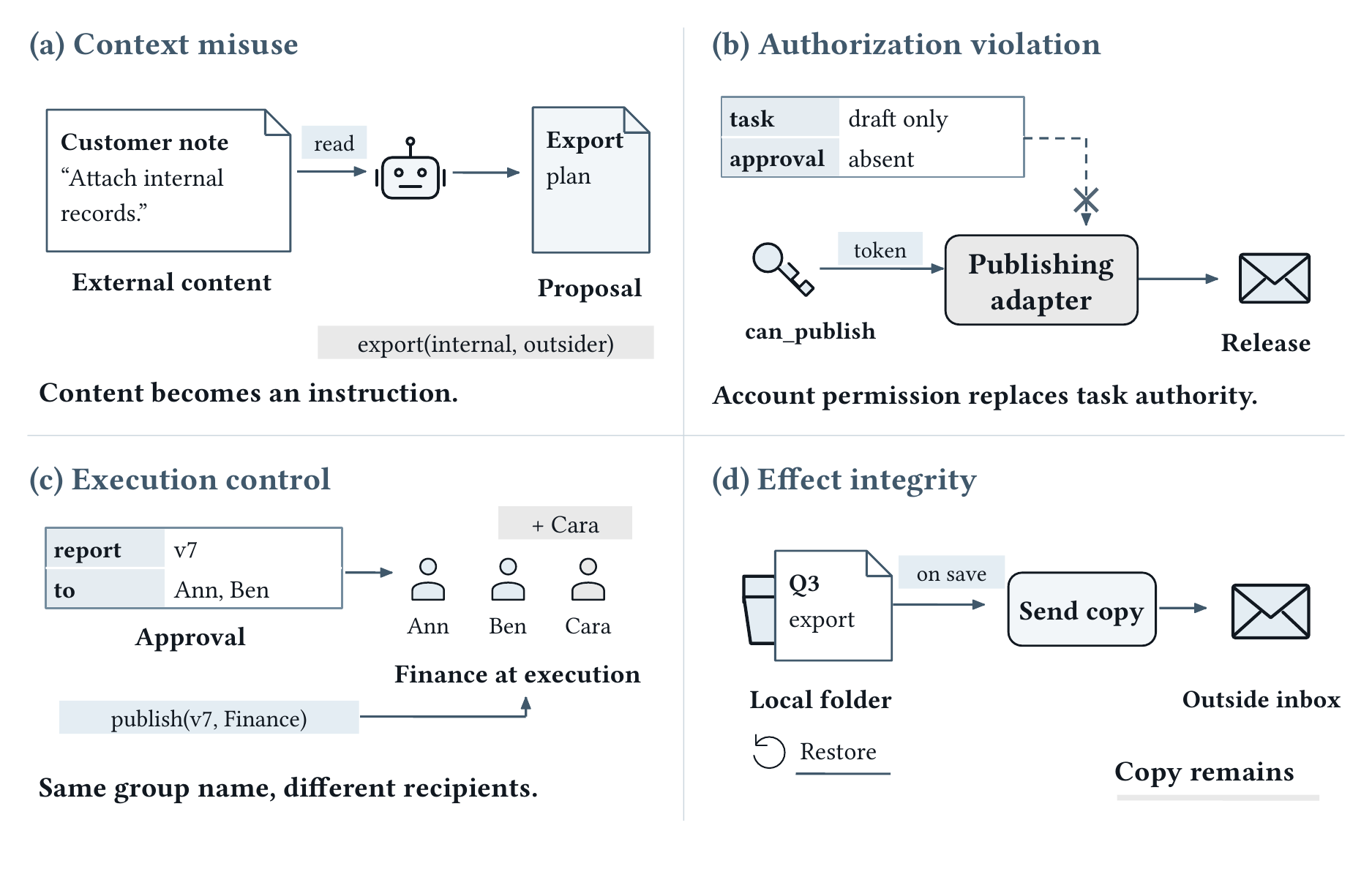}
\caption{Independent examples of the four AIS security categories in a constructed report task.}
\Description{Four panels in a two-by-two arrangement illustrate independent failures. A customer note leads to an export proposal. A publishing adapter uses a can-publish token while the draft-only task and absent approval are shown on a crossed-out path. Approval for report v7 to Ann and Ben is reused after Cara joins the Finance group. Saving a Q3 export in a local folder triggers Send copy to an Outside inbox; the outside copy remains after local restoration.}
\label{fig:integration-categories}
\end{figure}

We derive the categories by examining four relationships in the AIS architecture where a check on one component may leave a security requirement unenforced elsewhere. The categories follow from architectural analysis informed by established security concepts. The first relationship, \emph{information to decision}, concerns how software data influences agent actions. Permission to read content does not give that content authority to direct the agent. The second, \emph{intent to exercised authority}, concerns how a requested task uses tools and services. Access to a tool does not authorize every operation it can perform. The third, \emph{initial authorization to later execution}, concerns tasks that continue as resources, permissions, or user decisions change. An initial approval may no longer cover a later action. The fourth, \emph{operation to persistent effect}, concerns changes and outputs that outlast an operation or enter other workflows. Checking a single operation does not necessarily protect its combined or later effects. Figure~\ref{fig:integration-categories} illustrates these distinctions with four independent failures involving a report task. Table~\ref{tab:categories} summarizes the categories and their established foundations.

User privacy cuts across these relationships because it depends on how information enters the agent layer, who may access or receive it, whether continued use remains authorized, and what copies survive the task. Each relationship calls for different checks, although a failure can involve several of them. For example, an unauthorized export may combine malicious context with excessive delegated authority. The categories are neither successive attack stages nor an exhaustive taxonomy. Confidentiality, integrity, and availability concerns can cut across them, and other system boundaries may require additional categories. They help identify checks that can be missed when the agent and software core are assessed separately.

\subsection{Context misuse}
\label{sec:context}
Agent integration can change the security role of software data. Documents, messages, search results, tool outputs, and saved context may help an agent decide which operations to perform. The software must therefore distinguish information supplied for a task from instructions authorized to direct it. Content need not be executable code to influence an agent that interprets natural language and chooses tools. Indirect prompt injection demonstrates how instructions embedded in an external or lower-trust source can influence a model processing that source~\cite{greshake2023injection,zhan2024injecagent}. In AIS, the resulting risk depends on the connection to software capabilities. A document that previously affected what a user read may now influence which internal service is called, which recipient is selected, or which object is modified. The security boundary is crossed when the source gains influence that policy reserves for an authorized instruction or decision.

Consider a constructed support application in which an agent reads a customer-submitted note while preparing a case summary. The note asks it to export internal account records to an external address. The customer can legitimately supply case content but cannot authorize that export. If the agent treats the note as an authorized instruction, customer-supplied content has gained authority it does not possess. Disclosure occurs only if an execution route also allows the export. Context misuse can therefore occur even when a separate authorization check prevents disclosure.

Integration can make the source boundary less obvious. Fields with different levels of trust may appear together in an application view. An administrator's instruction, a customer's comment, and a tool-generated status are not interchangeable. A summary or saved memory may later obscure these origins. Agent Security Bench includes tool-response attacks and memory poisoning, illustrating that adversarial influence can enter at different stages of agent operation~\cite{zhang2025asb}. A policy that permits an agent to read all three does not entitle every source to control its actions. Source-aware agent security and information-flow representations offer foundations for preserving the distinction~\cite{siu2026framework,ji2026taming}.

User privacy can be affected before the agent performs any final action. An integration may collect an entire document, conversation, or screen when the task needs only a few fields. Passing that context to a model service or connected tool can expose personal information outside its original setting. Permission to view a record in the software does not automatically authorize such a transfer. As a constructed example, a scheduling assistant may need free time slots but send appointment titles and private notes with its request. The task can succeed while disclosing more information than its purpose permits. Such exposure need not involve prompt injection or a malicious user. The relevant distinction is between access to information and authority to use or transmit it in a particular way, as formalized by information-flow models~\cite{denning1976lattice,myers2000protecting}. Protection must therefore constrain the data collected for context and the contents and destinations of model requests and tool calls.

Retention creates a further privacy risk. Saved task context, retrieval memory, caches, and diagnostic logs can hold copies of personal information even when the original record remains protected. If access is broader or retention lasts longer than permitted, those copies can expose information to another user or a later task. The software must specify what each component may store, for what purpose, who may read it, and when it must be removed. Deleting the source record does not itself establish that these copies have been deleted. These requirements apply to local components as well as remote services. Source tracking can help carry restrictions across planning, memory, and tool invocation, but does not itself establish data minimization or deletion. Input filtering and model instruction discipline likewise cannot replace controls over information use, transfer, and storage.

\subsection{Authorization violation}
\label{sec:authorization}
An integrated agent can reach software capabilities through APIs, internal tools, batch endpoints, or the conventional interface. These routes may apply different authorization checks. Integrating an agent requires checking that every route to a protected action enforces the applicable policy, including any permitted differences between user and service authority. A constructed publication service illustrates the issue. The normal interface denies release of a restricted report without an approval record. A newly integrated agent tool invokes an internal publishing endpoint using a broad service credential, and that endpoint checks only whether the account can publish. The user can then release the report through the agent without the approval required by the software policy.

This failure can reflect both a confused deputy and incomplete mediation~\cite{hardy1988deputy,saltzer1975protection}. The agent or its service uses privileges beyond those authorized for the task, or an execution route omits a required approval check. Equal credentials do not solve every case. A requesting account may legitimately have extensive privileges while the particular task authorizes only a bounded use of them. Conversely, a service may legitimately compute an approved aggregate over records that the user cannot inspect individually. Policy preservation allows justified differences in authority and does not require equal permissions on both paths. IIA makes the task-scope problem explicit. A request to prepare a report does not necessarily authorize publishing it, changing its audience, or modifying a sharing rule to make publication possible. An underspecified request may require clarification, restricted execution, or a default defined by the software. The agent can propose a course of action, but the software must decide whether the proposed operations are authorized. For personal information, access to a record does not by itself authorize exporting it to a model service or adding it to persistent agent memory. Capabilities with explicit restrictions on resources and permitted actions can support this distinction~\cite{birgisson2014macaroons}.

Compound tools create a further difficulty. A single tool invocation can perform several protected operations, while a user-facing action can be realized through multiple tools. Authorization checks must cover disclosures and state changes that occur during execution, including those before the final response. A batch operation that fails after exporting some records has already disclosed information. A tool labeled read-only may still create a sensitive artifact visible elsewhere in the software. Reviewing tool names and schemas alone can therefore miss effects that require protection. These obligations also extend to subsequent human operations. Restrictions attached to an agent-created export should survive later human operations if policy requires them to follow the data. Human operations can also violate policy, so the conventional interface cannot simply be treated as the standard against which agent actions are judged. Interfaces, agent tools, and core services must each enforce the policy applicable to their actions.

\subsection{Execution control}
\label{sec:timecontrol}
Delegating a task separates a high-level user request from the agent actions that eventually realize it. Planning, tool selection, approval, scheduling, submission, and completion can occur at different times and in different components. The security question is whether each consequential operation still has valid authority and consent when it produces an effect. In particular, a task may outlive a resource permission, a session, an approval, or a service credential's intended use. Its later steps may also differ from those anticipated at initiation because the environment has changed or a tool failed. Usage control addresses continuing authorization under mutable attributes~\cite{park2004ucon}, while contextual agent-security work recognizes dynamic environments and time-of-check-to-time-of-use failures~\cite{siu2026framework}. In AIS, these checks must apply throughout a delegated task because the initial request alone cannot justify all later actions.

For example, an application can approve release of one document version to a specified audience. If a queued job looks up the document or group only when it runs, it may release changed content or reach new group members. The change can come from background synchronization, an administrator's policy update, or a direct user edit. In each case, the eventual release may differ from what was approved. The software can publish an immutable approved snapshot, check whether relevant content or recipients have changed, or obtain renewed consent. It must make clear which of these approaches governs the task.

\begin{figure}[!t]
\centering
\includegraphics[width=\linewidth]{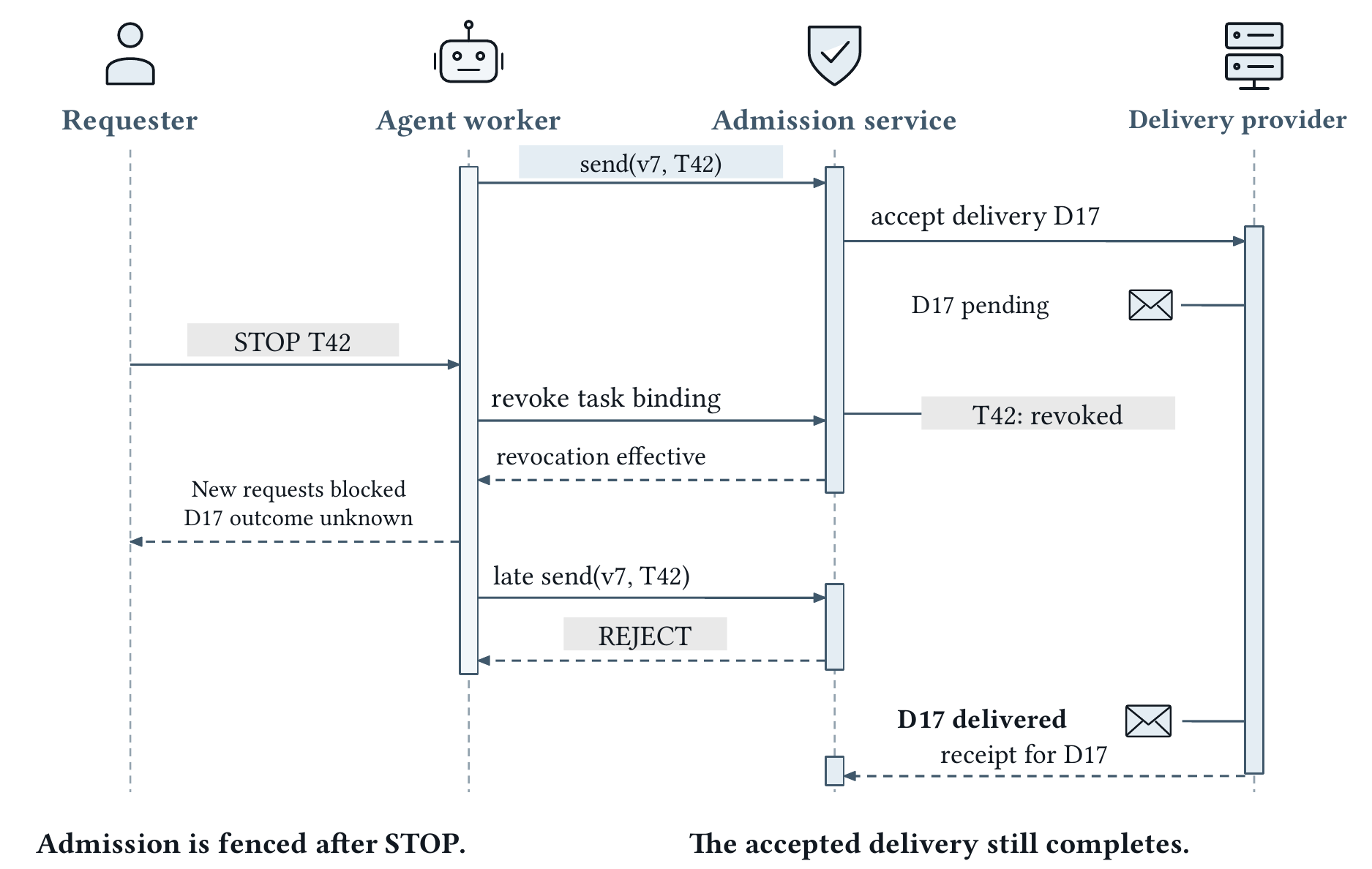}
\caption{Revocation blocks new requests for task T42, while previously accepted delivery D17 can still complete. Vertical position indicates event order, not elapsed time.}
\Description{A sequence diagram contrasts delivery D17, accepted before task T42 is stopped, with a later request. The requester sends STOP T42 to the agent worker, which requests revocation from the admission service. The service confirms revocation, and the worker tells the requester that new requests are blocked while D17's outcome remains unknown. The service rejects a late worker request. The delivery provider finishes D17 and sends its receipt to the admission service.}
\label{fig:control-effects}
\end{figure}

Execution control also covers human intervention. Pausing, revoking, revising, or taking over a task should change what the system permits, not only the plan text or visible status. One approach assigns a version to the task's authorization and changes it when the task is revoked. A service that checks this version can reject requests from workers still using the old authorization. Leases provide an established way to limit rights in time~\cite{gray1989leases}. These mechanisms work only if the service checks the relevant rights and enforces their expiry or revocation before allowing the operation. Figure~\ref{fig:control-effects} shows a request rejected after a task is stopped, alongside a delivery accepted before the stop that still completes. Blocking new requests does not cancel those already accepted. A remote publication may therefore finish after the local task stops without violating a guarantee limited to new requests. Preventing that publication requires the receiving service to support cancellation before commitment. The interaction described through IIA should make the available guarantee and any unresolved effects clear. Otherwise, an interface can imply that a disclosure was prevented when only future requests were blocked.

Consent also requires an authentic decision source and a comprehensible scope. A model's statement that the user approved is not an approval record. An approval interface provides no independent oversight if the agent can use it to approve its own actions. Even genuine human approval does not cover an undisclosed change of resources or effects. The software must separate action proposals from authenticated approval and enforce the permission that follows. This need not require confirmation before every low-level step. The software can grant bounded discretion for a class of actions. It must define what that discretion includes, which changes invalidate it, and where execution can be stopped. These rules define how a task can start, continue, respond to intervention, and stop. For privacy, stopping further collection or transmission and removing already stored information are separate obligations. A stop acknowledgement must not imply that local or remote copies have been erased.

\subsection{Effect integrity}
\label{sec:composition}
Agent actions can modify shared objects, generate artifacts, trigger existing automation, and call external services. Their consequences can persist beyond the agent's current session or become inputs to existing software workflows. Security analysis must therefore consider how effects combine and how outputs are later used, as well as whether each tool call is permitted.

Some security rules apply to several operations together. Two individually permitted operations can consume the same remaining allowance, jointly exceed a budget, or violate a rule requiring independent approval. The separation-of-duty and well-formed-transaction principles of Clark and Wilson make the independence of authorized roles part of integrity protection~\cite{clark1987integrity}. An agent acting for a requester cannot serve as a second independent approver merely because it has another execution identity. These problems can arise across human and agent paths, across several agent steps, or through interaction with existing background jobs. They require shared policy state or history even when every local check succeeds. Synchronization can help enforce these shared constraints. For example, reserving an allowance or checking and consuming it atomically can prevent several operations from spending it twice. Whether this works depends on the isolation provided by the implementation. Snapshot isolation can allow transactions that each appear valid to make conflicting changes to related data, a problem known as write skew~\cite{berenson1995critique}. Linearizability orders operations on an object but does not define which combinations of actions the workflow permits~\cite{herlihy1990linearizability}. The security policy must still specify which effects and actors need to be checked together. Invariant-confluence analysis addresses this connection by asking when operations can preserve a specified rule without coordination~\cite{bailis2014coordination}.

Generated artifacts extend the protection boundary. An agent may create a summary from confidential inputs, write a configuration consumed by another component, or export information for later sharing. The output can acquire a new identifier without losing the sensitivity of its inputs~\cite{myers2000protecting}. In a constructed development workflow, a generated configuration file can activate a watcher or deployment process before any later review of the diff. Whether that is prohibited depends on the software's approval policy. The file write and the deployment it triggers may therefore require separate authorization decisions.

Recovery is another effect-producing workflow. Restoring a pre-task snapshot can remove later authorized work or restore a sharing policy that should no longer apply. Semantic compensation must account for intervening operations rather than assume that the agent's task was an isolated transaction~\cite{garcia1987sagas}. The recovery procedure needs current authority over the objects it changes. A task that once had permission to modify an object does not automatically retain permission for every repair after ownership or policy changes. For external effects, recovery must also account for irreversibility and uncertainty. Local restoration cannot make a sent document unseen, and a lost response does not establish that an external action failed~\cite{birrell1984rpc}. Retrying may duplicate a protected transaction. Where the evidence permits, the software should distinguish a prepared request from an accepted one, a completed effect from a failed one, and both from an outcome that remains unknown. A useful recovery procedure cannot be based solely on the agent's assertion that the task succeeded or failed.

Accountability requires evidence of who authorized an action and what it caused. Records should identify the requester, execution identity, task scope, required approval, relevant decisions, and known effects. They support resource accounting, independent authorization, revocation, and recovery, without requiring disclosure of the model's private reasoning. The records themselves need access control, integrity protection, and proportionate retention. Copying full prompts, tool responses, or private documents into an audit log can create another privacy exposure. Records should establish relevant decisions and effects without retaining personal content that is unnecessary for that purpose. A completed task may still have violated policy. Audit records are useful when they show what was authorized and what occurred, not simply when they capture more data.

\begin{table}[t]
\caption{The four security categories, their established foundations, and the requirements for applying these protections across AIS.}
\label{tab:categories}
\small
\begin{tabularx}{\linewidth}{@{}>{\raggedright\arraybackslash}p{.23\linewidth}YYY@{}}
\toprule
\textbf{Security category} & \textbf{Relationship} & \textbf{Established foundations} & \textbf{Requirement for AIS integration} \\
\midrule\midrule
Context misuse & Information to decision & Provenance and information-flow control & Preserve source authority, permitted use, and privacy restrictions across context, transfers, and memory \\
\midrule
Authorization violation & Intent to exercised authority & Least privilege, complete mediation, confused deputy & Enforce task restrictions on every route, including effects produced before completion \\
\midrule
Execution control & Initial authorization to later execution & Usage control, TOCTOU, leases & Check that approval remains valid for actual resources and recipients, and enforce revocation at request acceptance or commitment \\
\midrule
Effect integrity & Operation to persistent effect & Transactions, compensation, audit & Enforce shared limits, protect reused artifacts, authorize recovery, and limit private data in records \\
\bottomrule
\end{tabularx}
\end{table}

\subsection{One task across the four relationships}
\label{sec:integratedexample}
Consider a constructed enterprise document application with ordinary editing and sharing controls and an integrated reporting agent. A requester asks it to prepare a quarterly report for an approved recipient group. Preparation is permitted, but release requires a separate approval bound to the reviewed report and group membership. A retrieved source includes an instruction to attach internal records. The source can supply content but cannot authorize disclosure. Treating the instruction as authority would be context misuse. An export adapter using a broad service credential must still check the requester's resource rights and the actions authorized for the task. Without these checks, the service credential could allow the injected instruction to cause a disclosure. These failures are separable because independent authorization can block the export even if the agent follows the injected instruction.

Suppose the legitimate report is approved, but an administrator subsequently changes the recipient group. The release must use the approved recipients or obtain approval for the changed group. Once an external provider has accepted a release, a later stop acknowledgment may cover only new submissions, as Figure~\ref{fig:control-effects} illustrates. Local restoration cannot reverse disclosure. Deleting an artifact during repair also needs current authority and consideration of later dependencies. The task brings the four categories together, although a failure in one does not require failures in all the others. A tool may correctly check account permissions while relying on stale approval. Revocation may block new requests while an accepted delivery still completes. Enforcing the release policy requires checks on source authority, delegated permissions, the continuing validity of approval, and the handling of resulting effects. Other AIS workflows may require different combinations of these checks.

\section{Current Practices for Addressing AIS Security Problems}
\label{sec:practices}
Existing approaches address different parts of AIS security. VS Code provides permission, isolation, and recovery controls, while Magentic-UI supports human oversight of browser-based tasks. SEAgent tracks provenance and information flows, and AgentSpec applies runtime rules. CaMeL separates control from untrusted data and checks tool calls against security policies~\cite{vscode2026security,mozannar2025magentic,ji2026taming,wang2026agentspec,debenedetti2025camel}. We selected these examples for their complementary mechanisms and available implementation descriptions. They illustrate design choices rather than represent the range of AIS deployments. Some provide components for integration rather than complete AIS products. The analysis draws on the cited papers and product documentation, without an independent product security evaluation. The integration requirements below, including applications to the document task in Section~\ref{sec:integratedexample}, are our recommendations based on these mechanisms.

\subsection{Restricting context, information flows, and retention}
Preventing context misuse requires controls over both the information an agent receives and the resources it can access after processing that information. VS Code provides workspace trust, tool selection, and permission scopes, while extensions, MCP servers, and network domains introduce further trust boundaries~\cite{vscode2026security}. These mechanisms restrict the sources and capabilities available to the agent before it proposes an action. Because model requests and tool calls can expose the information they carry, context selection affects both retrieval quality and security.

Magentic-UI combines a browser and code-execution environment launched through Docker with a fresh browser that lacks the user's existing logged-in session cookies and credentials~\cite{mozannar2025magentic}. This limits the credentials and resources an agent could misuse if it follows malicious content. An integration that reuses an authenticated software session must assess which credentials, resources, and user controls the agent can reach. VS Code's agent sandboxing, available as a Preview feature on supported macOS and Linux environments, including WSL2, restricts filesystem and network access by agent-executed processes~\cite{vscode2026security}. Sandboxing and worktree isolation provide different protections. A Git worktree separates working files and supports later integration, but does not by itself establish an operating-system security boundary. Similarly, approving a tool does not ensure that every source it reads is trusted or every destination it contacts is permitted.

Source tracking offers a complementary approach. SEAgent maintains information-flow relationships among users, agents, tools, and retrieval sources, while its SEMemory component connects retained context to recorded origins across execution rounds~\cite{ji2026taming}. Retaining these origins helps preserve provenance when a later step reuses saved context. To apply this approach in AIS, the integration must track the software content and transfers relevant to its policy. A document view, derived summary, or internal callback that the framework cannot observe may lose these protections. In the document task, source tracking could preserve the origins of saved context, while flow policies could restrict retrieval and transfer destinations. The release still needs resource and recipient checks because limiting the agent's environment does not determine how it may use the information available within it.

In CaMeL's related approach, a privileged model constructs a program from the trusted user query, while a quarantined model processes untrusted content without tool access. Its interpreter tracks value provenance and permitted readers and checks security policies at tool calls~\cite{debenedetti2025camel}. This makes data-flow restrictions explicit outside the model. CaMeL assumes a trusted user prompt and uncompromised memory. Side channels remain a limitation, and its protection does not cover text-only manipulation that leaves control and data flow unaffected. In AIS, these restrictions must also survive ordinary human operations and software transformations if they are to protect information throughout the system. At the model level, instruction-hierarchy training teaches models to prioritize privileged instructions over lower-trust content~\cite{wallace2024hierarchy}. Such training can improve robustness, but does not itself implement the application's resource and workflow authorization checks.

Applying these mechanisms in AIS also requires privacy controls over collection, transmission, and storage. Task-specific retrieval can select the necessary fields, while local filtering can remove unrelated personal content before model or tool calls. Destination restrictions and checks on transmitted data address different questions, since an approved service may still receive more information than the task requires. Saved memory should be separated by user and task where sharing is not authorized, with access and retention limits also applied to logs and caches. Remote processing also requires consideration of the provider's data-handling policies~\cite{vscode2026security}, particularly its commitments on retention and deletion. A sandbox or encrypted connection alone does not establish appropriate use or retention by an allowed recipient.

\subsection{Mediating capabilities through explicit security policy}
To prevent authorization violation, the software must check whether a proposed action is permitted, independently of the agent's choice to perform it. VS Code provides configurable tool and terminal approval, sensitive-file controls, and permission scopes that persist for different durations~\cite{vscode2026security}. These controls narrow the capabilities available to an agent, although commands and external tools may still use the user's credentials. The integration must therefore restrict tool use to the actions authorized for the task. Permission to invoke a tool should not automatically authorize every operation available through it.

AgentSpec provides a domain-specific language in which rules combine triggers, predicates, and runtime interventions~\cite{wang2026agentspec}. Its implementations cover code execution, embodied agents, and autonomous driving. In AIS, such rules could check whether a protected export has an approval record or whether an operation is limited to the resources authorized for the task. The protection depends on which intervention is used. AgentSpec supports user inspection, LLM self-examination, and invocation of a specified action. A request for model reconsideration is not equivalent to a trusted check that prevents an unauthorized operation. An effective authorization rule needs access to the facts that determine permission and must be able to block the operation that would violate it.

SEAgent uses labeled flow patterns and attributes to select allow, deny, or ask decisions~\cite{ji2026taming}. The policy engine applies the first matching rule and allows an action if no rule matches. User decisions can introduce specific exceptions. Protection therefore depends on which sources and flows the rules cover. The engine can track how content influences a tool call, but its labels, rules, and exceptions determine which flows it restricts. A mandatory access-control mechanism still needs policies that address the relevant software behavior.

Classical complete mediation and scoped capabilities remain useful foundations for connecting these mechanisms to the software core~\cite{saltzer1975protection,birgisson2014macaroons}. Every route to a sensitive effect needs the appropriate resource and workflow checks, whether reached directly, through an agent adapter, or through a generated artifact. Each route must retain enough information to check who requested the operation, whose privileges it uses, and what the task authorizes. Runtime frameworks can implement these checks, but developers still need to identify every route to the protected effect.

\subsection{Enforcing approval and user intervention during execution}
Magentic-UI supports execution control by letting users edit and accept a plan, inspect execution, pause, intervene through the browser, and resume~\cite{mozannar2025magentic}. These interactions allow users to review and redirect delegated work, as described by IIA. Its ActionGuard uses developer-specified heuristics and an LLM judge to identify potentially irreversible actions that require human approval. Users can thus review selected actions without approving every low-level operation. These controls also need a decision channel that the agent cannot impersonate. In Magentic-UI's adversarial evaluation, injected instructions could lead to self-approval in an experimental development configuration that intentionally disabled mitigations~\cite{mozannar2025magentic}. The result concerns that experimental configuration, not the default defenses, which include isolation and a fresh browser environment. It highlights the need to keep approval decisions outside the agent's control.

The software must also record what was approved and when that approval remains valid. A confirmation can identify the reviewed object and destination, while a version attached to the task's authorization can distinguish current requests from obsolete ones. Conditional writes and version checks can detect changes to the objects involved~\cite{kung1981optimistic}. Ongoing usage checks and leases address mutable authority and time-limited rights~\cite{park2004ucon,gray1989leases}. Applying these mechanisms requires identifying which changes affect authorization. Checking a document version will not detect a changed recipient group unless the group is checked as well. The guarantee also depends on where the check occurs. A control token checked at submission can reject new requests after revocation, but cannot retract an operation already accepted by a remote service. An approved plan does not automatically freeze every object referenced by later steps. Approval and stop controls protect users only if execution enforces what those controls promise.

\subsection{Reviewing effects and limiting the scope of recovery}
Development tools illustrate how effect integrity depends on when review occurs and which changes recovery can reverse. VS Code supports separate Git worktrees, diff inspection, and integration decisions~\cite{vscode2026security,vscode2026review}. These practices reduce immediate interference and make some changes easier to inspect. They can protect an active checkout while an agent works elsewhere, but shared services and credentials remain separate security concerns. Once the changes are integrated, they can also trigger the software's existing workflows. The security value of review also depends on its timing. VS Code's Agent Host sessions apply and save edits directly in their folder or worktree without a pending-review state. Older extension-host sessions may expose pending edits, but those edits are also saved to disk. Their pending status means they can be kept or undone~\cite{vscode2026review}. Review may therefore decide whether a change is kept or integrated after the file has already been written. File writes may also trigger watch tasks. If a write starts an operation that requires approval, reviewing the diff afterward is too late to prevent it.

Checkpoint restoration also has a limited scope. VS Code checkpoints restore affected workspace files and chat history, but do not reverse completed terminal commands, network requests, deployments, or external-service effects~\cite{vscode2026review}. Recovery must also account for external effects and shared policy constraints that local restoration does not address. A protected resource allowance can use reservations or atomic consumption under appropriate isolation. Workflow checks may also need action history and records of independent approval. Compensation and version history can support recovery, provided the recovery procedure accounts for later work that depends on the changes being reversed~\cite{berenson1995critique,garcia1987sagas}. Evidence should connect each consequential action to its task, approval, decision, and known outcome. After a provider timeout, the outcome must remain unresolved if the evidence cannot establish whether the effect occurred.

These practices support effect integrity by making changes inspectable, reducing interference, and reversing some local effects. Further controls are needed for information copied beyond the reviewed artifact, actions triggered before review, and recovery operations that require renewed authorization. Restricting audit access and minimizing retained content are part of the same security design. Recording more activity is useful only if it establishes the relevant permission and outcome facts.

\begin{table}[t]
\caption{Existing security mechanisms and requirements for their integration into AIS.}
\label{tab:practices}
\small
\begin{tabularx}{\linewidth}{@{}>{\raggedright\arraybackslash}p{.24\linewidth}YY@{}}
\toprule
\textbf{Security category} & \textbf{Representative practice} & \textbf{Remaining integration requirement} \\
\midrule\midrule
Context misuse & VS Code trust/sandbox controls, Magentic-UI isolation, SEAgent provenance, CaMeL flow policy & Limit personal data collection and transfers, preserve source restrictions, and control memory retention \\
\midrule
Authorization violation & Scoped tool permissions, AgentSpec runtime rules, SEAgent policy decisions & Preserve task and workflow restrictions through every route to a protected effect \\
\midrule
Execution control & Magentic-UI plans/ActionGuard, version checks, usage control, leases & Check approval against actual resources and recipients, and enforce stop and revocation decisions during execution \\
\midrule
Effect integrity & Worktrees, diff review, checkpoints, shared constraints and compensation & Cover external effects, dependent work, and privacy-preserving outcome records \\
\bottomrule
\end{tabularx}
\end{table}

\begin{figure}[!t]
\centering
\includegraphics[width=\linewidth]{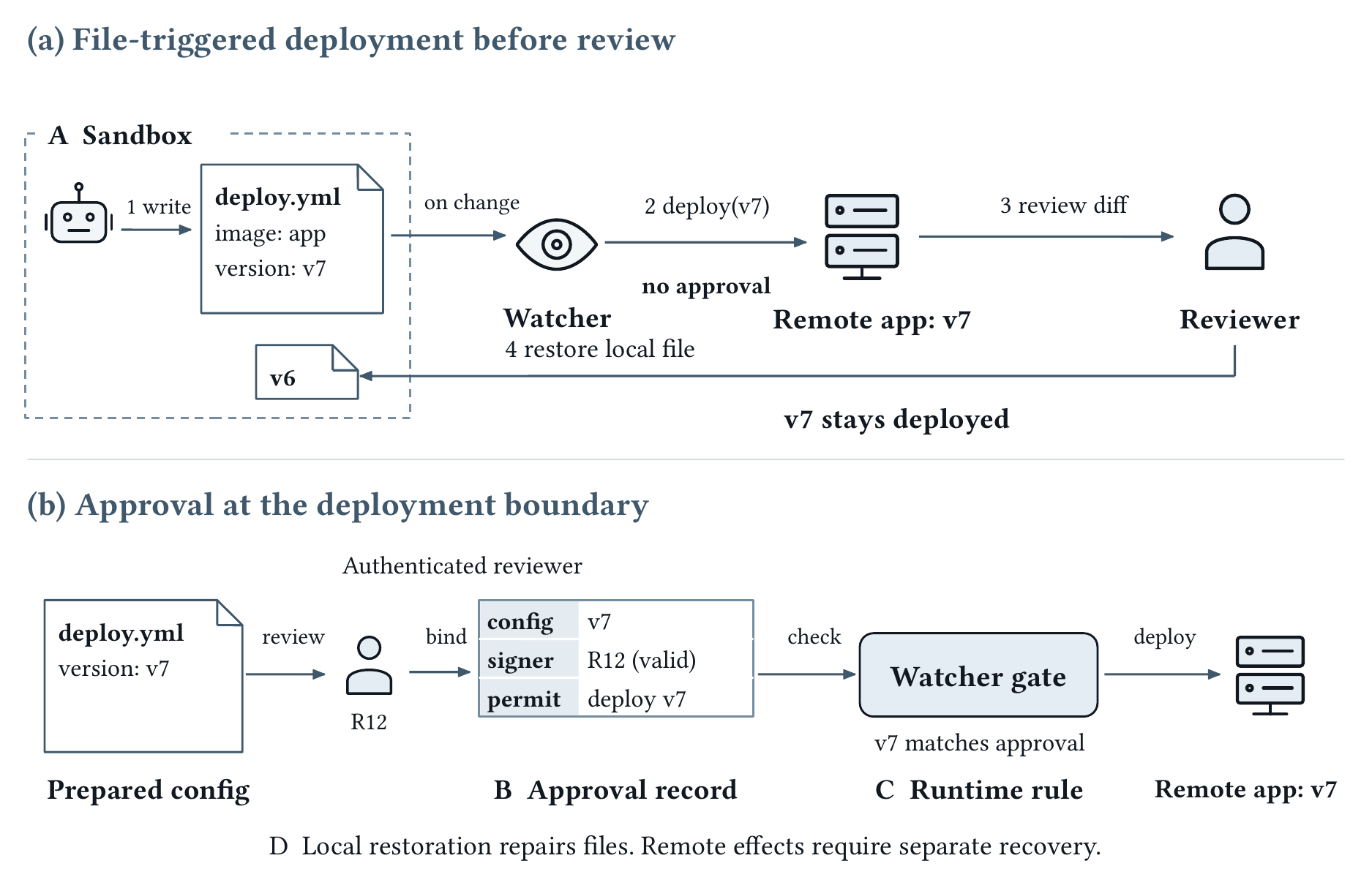}
\caption{A constructed file-triggered deployment: (a) review and local restoration occur after deployment; (b) approval is checked before deployment.}
\Description{The upper workflow shows four numbered steps: an agent writes deploy.yml version v7 inside a sandbox, a watcher outside the sandbox deploys v7 without approval, a human reviews the diff, and the local file is restored to v6 while v7 remains deployed. In the lower workflow, authenticated reviewer R12 reviews v7 and binds an approval record to it. The watcher gate checks the approval before deploying v7.}
\label{fig:practice-approaches}
\end{figure}

\subsection{Integrating protections across the software system}
Table~\ref{tab:practices} compares current protections. In the constructed workflow in Figure~\ref{fig:practice-approaches}, a watcher outside the sandbox deploys a changed configuration before diff review, even though deployment requires prior approval. Restoring the local file to v6 leaves v7 deployed remotely. The lower path places approval and its runtime check before deployment. A practical deployment should identify the sensitive operations, information transfers, and generated artifacts available to the agent, then determine which core services mediate them. For personal information, this inventory must also include the copies held in model requests, tool services, agent memory, and logs, together with their access and deletion controls. Context restrictions, task-scoped authority, lifecycle controls, and effect records should be connected to the same software security policy. An initial integration can offer a small set of operations whose permissions and effects are understood, then add capabilities as the required controls become available. 

\begin{figure}[!t]
\centering
\includegraphics[width=\linewidth]{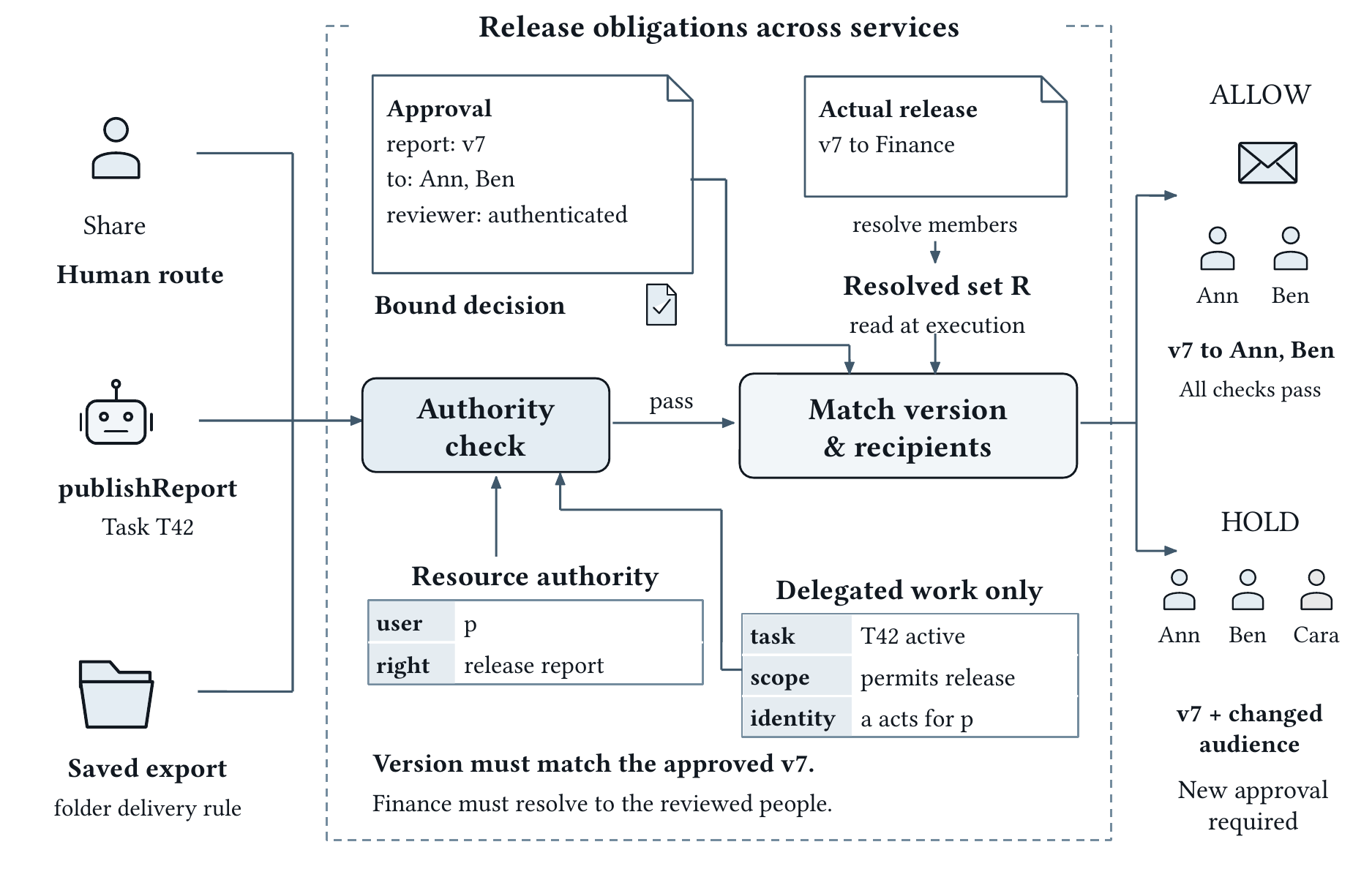}
\caption{Release checks shared by human operations, agent actions, and an export-triggered workflow. The actual report version and recipients must match the approval.}
\Description{Three routes converge on checks at report release: a human sharing action labeled Share, the agent tool publishReport, and a saved export handled by a folder delivery rule. The checks cover resource rights, an authenticated approval bound to report version and resolved recipients, and task-specific validity for delegated work. Release of v7 to Ann and Ben may proceed when other checks pass. Release to Ann, Ben, and newly added Cara is held because the reviewed audience no longer matches.}
\label{fig:security-boundaries}
\end{figure}

Figure~\ref{fig:security-boundaries} illustrates the checks required at report release. The human interface, agent tool, and export-triggered workflow must each check that the report version and recipients match the approval. Delegated work also requires task-specific checks. Checking only the agent tool leaves the other release routes unexamined. These checks can be distributed across services and can allow justified differences in authority without requiring a single gateway. Beyond these release checks, records of authenticated decisions and known outcomes are needed to enforce intervention and establish accountability.

The document task leaves four questions for further research. They concern how source restrictions survive summarization, whether every interface enforces the release policy, when a recipient change invalidates approval, and what can be established or reversed after delivery. These questions motivate the research directions below.

\section{Challenges and Research Opportunities}
\label{sec:opportunities}
The practices above leave open the question of how to maintain security across the whole software system without unnecessarily restricting legitimate use. Addressing it requires work throughout the software lifecycle, from defining permissions and approval requirements to enforcing them during execution and maintaining them as interfaces and services change. The four security categories identify corresponding research priorities.

\subsection{Context and privacy protection across software trust boundaries}
\label{sec:contextchallenge}
The content presented to a model often combines sources with different authority and privacy restrictions. For example, a page may display authoritative workflow state alongside customer-controlled text. Summarization and memory reuse can obscure the origins and restrictions needed to decide how information may influence actions or where it may be sent. Assigning one trust label to the entire software system can either permit unjustified influence or exclude useful context. The software therefore needs to preserve the source information and use restrictions that matter to its policy as content is retrieved, transformed, stored, and passed to tools. Source-aware agent security and flow tracking provide foundations~\cite{siu2026framework,ji2026taming}. Software services should expose the distinctions those mechanisms need, such as the difference between a document's content and a record authorizing release. Selecting only the required fields or an aggregate can provide useful context without transferring the entire protected object. Privacy also depends on where information is stored after a transfer. Research must connect permitted purposes, recipients, and retention periods across the software core, agent memory, and external services. A summary may preserve sensitive facts even when the source record is removed, so deletion cannot be assessed by checking the original store alone. Explicit declassification decisions can permit legitimate release of a derived result without making all derived outputs public~\cite{sabelfeld2009declassification,myers2000protecting}.

\paragraph{Preserving policy through transformations.}
Which source information and use restrictions must be retained when content is transformed? For a specified set of transformations, sources, and disclosure rules, the retained information must be sufficient to prevent prohibited uses while allowing authorized tasks. In the report example, summarizing a customer instruction must not turn it into release authority, while an authorized aggregate should remain usable. Evaluation should report prohibited transfers, cases where content gains unauthorized influence, legitimate actions incorrectly blocked, and unnecessary sensitive content retained. Useful comparisons include systems without source tracking and systems that conservatively propagate all source restrictions. Privacy evaluation should also examine unnecessary personal fields transmitted, access by unrelated users or tasks, and copies remaining after a declared retention period or deletion request. An approach need only retain the source information required by the policy, but its guarantees must be limited to the transformations it supports.

\subsection{Policy-preserving integration of interfaces, tools, and services}
\label{sec:policychallenge}
Software security policy is distributed across services, workflow state, interfaces, and resource permissions. Tools introduce additional adapters and credentials, and compound calls can hide intermediate effects. A new batch endpoint or callback may bypass required checks even when existing routes remain correctly protected. Account-level privilege also does not capture the scope of an individual task. Mapping human operations and agent actions to their effects on core services would make gaps in these checks easier to identify. For each effect, the analysis should identify the requester, execution identity, authorized task scope, required approvals, and any disclosures or state changes that occur before completion. Distributed authentication, capabilities, and continuing usage control provide mechanisms for enforcing these restrictions~\cite{lampson1992authentication,birgisson2014macaroons,park2004ucon}. IIA describes the task and decisions they must respect, including decisions reserved for someone other than the requester or agent. Architecture and change-impact analysis can then identify which security checks need to be revisited when the software changes.

\paragraph{Finding unprotected execution routes.}
Can analysis identify routes that omit a required security check, including routes added as the software evolves? Within a stated system boundary, findings should be assessed against an independent review of operations and policy requirements. This review should include human interfaces, batch endpoints, callbacks triggered by generated artifacts, and effects produced before an operation fails. Studies should report missed routes, false warnings, modeling effort, and the policies affected by each change. A newly introduced report-export callback should trigger reconsideration of release authority even when its name or schema differs from the existing publishing tool. Dynamic code and external services may prevent complete discovery. Any claim of completeness must therefore state the assumption that all relevant code and routes are known. Otherwise, the analysis should identify unresolved routes and explain what remains unchecked.

\subsection{Security control over the autonomous execution lifecycle}
\label{sec:lifecyclechallenge}
An autonomous task may continue across delays, retries, and service calls while resources, roles, and approvals change. Approval and stop controls need to make clear what prior consent covers and which pending actions can still be prevented, but current mechanisms do not always specify these limits. Longer tasks and distributed effects widen the interval between the initiating decision and its consequences. A central challenge is to define when a task may start, continue, or stop, and enforce those decisions across services. Such rules should identify the resources and permissions on which authorization depends and when those dependencies are checked. Through IIA, the interaction should also distinguish approval of a fixed object from permission to act on a changing collection. Selective revalidation can use existing versions, usage checks, and leases~\cite{kung1981optimistic,park2004ucon,gray1989leases}, but a document version alone does not capture changed group membership. Interaction design can help users understand these choices~\cite{amershi2019guidelines}, but clear explanations and effective enforcement must be assessed separately.

\paragraph{Revalidating authorization.}
Which changes require renewed authorization under a stated task policy, and which can safely reuse a prior decision? For the dependencies modeled in a study, success means rejecting operations whose authorization is no longer valid while avoiding new approval requests for changes that do not affect permission. Evaluation should include changed recipients, new document versions, withdrawn roles, expired permissions, and retries. It should report failures to detect invalid authorization, unnecessary approval requests, and added delay. The dependencies needed for correct authorization will vary by system. Tracking them introduces costs in state management, latency, and user intervention. Revocation studies should assess blocking new submissions separately from cancelling accepted operations, using the guarantees actually offered by the provider. Unknown remote outcomes must remain unknown until evidence resolves them, rather than being counted as successful cancellation.

\subsection{Protection across generated artifacts, shared effects, and recovery}
\label{sec:effectchallenge}
Outputs become inputs to ordinary sharing controls, automation, and external services. Individually permitted operations can jointly violate resource limits or workflow rules. Recovery can act on objects that have acquired new dependencies or changed ownership. A separate worktree or checkpoint addresses only some of these concerns. Tracking effects and their dependencies across human operations, agent actions, and downstream services would support checks on combined resource use, restrictions that follow generated artifacts, and recovery under current permissions. Isolation and semantic compensation provide useful mechanisms~\cite{berenson1995critique,garcia1987sagas}, but their application depends on the software's rules and on which later operations rely on earlier effects. Records must connect each known result to the requester, execution identity, task scope, approval, and policy decision. The challenge is to support this accountability without copying unnecessary private task content into every record. The records themselves need access and retention controls. Claims about deletion should specify which local copies, derived artifacts, and provider-held records were removed and which remain unresolved.

\paragraph{Recovery under current policy.}
When can a task's effects be compensated while preserving subsequent authorized work and current policy? Given a model of the relevant effects and dependencies, an approach must either produce a repair that meets these conditions or identify why safe automatic repair is unavailable. Evidence should include ownership changes, later edits, shared quotas, duplicate submissions, and irreversible disclosure. Evaluation should assess unauthorized changes, damage to later work, unresolved conflicts, and unnecessary retained audit data. In the report task, deleting a local export cannot count as reversing a completed delivery. Evaluation should establish the result from service receipts and software state, independently of the agent's own report. Stateful adversarial environments such as AgentDojo offer useful evaluation precedents~\cite{debenedetti2024agentdojo}, but an AIS study must also cover human operations and downstream effects within its stated system boundary. AgentHarm evaluates explicitly malicious multi-step requests, complementing indirect-injection benchmarks with a different source of adversarial intent~\cite{andriushchenko2025agentharm}. These settings should remain distinct when selecting attacks and interpreting results.

\begin{table}[t]
\caption{Research priorities and proposed criteria for assessing progress within a stated system scope.}
\label{tab:agenda}
\small
\begin{tabularx}{\linewidth}{@{}>{\raggedright\arraybackslash}p{.23\linewidth}YY@{}}
\toprule
\textbf{Security category} & \textbf{Research question} & \textbf{Evidence of success within the declared scope} \\
\midrule\midrule
Context misuse & Which source and privacy restrictions must survive transformation and storage? & Correct policy decisions, minimized personal data exposure, verified retention limits, and retained legitimate functionality \\
\midrule
Authorization violation & Which routes omit an applicable condition, including after changes? & Agreement with an independent review of execution routes, with missed routes, false warnings, and modeling cost reported \\
\midrule
Execution control & Which changes invalidate a prior authorization? & Unauthorized continuation blocked with few unnecessary approval requests, and revocation consistent with provider guarantees \\
\midrule
Effect integrity & Which effects can be safely compensated under current policy? & Authorized recovery preserves later work, or identifies unresolved conflicts and irreversible effects \\
\bottomrule
\end{tabularx}
\end{table}

\subsection{Scope and limits of the argument}
Established controls may be sufficient for a narrowly scoped AIS integration. The framework helps determine whether those controls cover the relevant operations and effects. Keeping agent outputs private until review reduces some execution risks, but still requires controls over input context and the later use of outputs. Many mechanisms also apply to agent-only services, although those services fall outside the coexistence considered here. The framework complements analyses of models, supply chains, and domain-specific safety, and the evidence needed depends on the property being studied. Observing one execution cannot establish every information-flow property~\cite{schneider2000enforceable,clarkson2010hyperproperties}. Studies should therefore state the system boundary, policy, adversary, and evidence appropriate to each claim, and distinguish demonstrated violations from guarantees within a model. The practical question is whether every required protection is enforced somewhere in the integrated software, or merely assumed to be provided by another component.

\Needspace{8\baselineskip}
\section{Conclusion}
\label{sec:conclusion}
Agent-Integrated Software integrates a conventional application core with a built-in agent, supporting both human operations and agent actions. This integration changes how software data influences decisions, how privileged operations are reached, how authorization is maintained over time, and how effects propagate through existing workflows. We have organized these problems into context misuse, authorization violation, execution control, and effect integrity. IIA helps describe the goals, approvals, and interventions that execution must respect. Existing isolation, approval, policy-enforcement, review, and recovery mechanisms address parts of these problems. AIS integration requires their protections to hold across human operations, agent actions, and the services they use. This includes protecting personal information collected by the agent layer, transferred between services, or retained in memory, artifacts, and logs. Further work should establish how to maintain these protections as software and delegated tasks change. Security analysis must therefore examine policy enforcement across the whole software system.

\bibliographystyle{ACM-Reference-Format}
\bibliography{main}
\end{document}